\documentclass[superscriptaddress,onecolumn,10pt,pra,aps,showpacs,longbibliography, notitlepage]{revtex4-1}
\usepackage{float}
\floatstyle{boxed}
\usepackage{wrapfig}
\usepackage{array}
\usepackage{graphicx}
\usepackage{amsbsy}
\usepackage{epstopdf}
\usepackage{amsmath,amssymb,amsfonts}
\usepackage{dsfont}

\usepackage[dvipsnames]{xcolor}

\def \am{\hat a}
\def \ap{\hat a^{\dagger}}

\newcommand{\de}{{\rm d}}

\newcommand{\bra}[1]{\langle #1|}
\newcommand{\ket}[1]{|#1\rangle}

\newcommand{\expec}[1]{\left\langle #1 \right\rangle}

\newcommand{\Tr}[1]{\text{Tr}\left[ #1 \right]}

\renewcommand{\eqref}[1]{\mbox{Eq.~(\ref{#1})}}
\renewcommand{\Re}[1]{{\rm Re}\left[#1 \right]}

\newcommand{\be}{\begin{equation}}
\newcommand{\ee}{\end{equation}}
\newcommand{\bea}{\begin{eqnarray}}
\newcommand{\eea}{\end{eqnarray}}

\usepackage{multirow}

\usepackage{url}
\usepackage[colorlinks]{hyperref}
\hypersetup{
    plainpages=true,
    breaklinks=true,
    hypertexnames=false,
    pageanchor=true,
    colorlinks=true,
    linkcolor={blue},
    citecolor={red},
    urlcolor={blue},
    anchorcolor={black}
}

\newcommand{\LL}{\mathcal{L}}
\newcommand{\DD}{\mathcal{D}}
\newcommand{\rhot}{\hat{\rho}(t)}
\newcommand{\sss}{\hat{\rho}_{\rm ss}}
\newcommand{\eig}[1]{\hat{\rho}_{#1}}

\makeatletter
\newcommand*\bigcdot{\mathpalette\bigcdot@{.5}}
\newcommand*\bigcdot@[2]{\mathbin{\vcenter{\hbox{\scalebox{#2}{$\m@th#1\bullet$}}}}}
\makeatother

\usepackage{enumerate}
\usepackage[normalem]{ulem} 

\begin{document}

\author{Fabrizio Minganti }
\email{fabrizio.minganti@gmail.com} 
\affiliation{Theoretical
Quantum Physics Laboratory, RIKEN Cluster for Pioneering Research,
Wako-shi, Saitama 351-0198, Japan}
\affiliation{Institute of Physics, Ecole Polytechnique F\'ed\'erale de Lausanne (EPFL), CH-1015 Lausanne, Switzerland}
\author{Ievgen I. Arkhipov}
\email{ievgen.arkhipov@upol.cz} \affiliation{Joint Laboratory of
Optics of Palack\'y University and Institute of Physics of CAS,
Faculty of Science, Palack\'y University, 17. listopadu 12, 771 46
Olomouc, Czech Republic}
\author{Adam Miranowicz}
\email{miran@amu.edu.pl} \affiliation{Theoretical Quantum Physics
Laboratory, RIKEN Cluster for Pioneering Research, Wako-shi,
Saitama 351-0198, Japan} \affiliation{Institute of Spintronics and
Quantum Information, Faculty of Physics, Adam Mickiewicz
University, 61-614 Pozna\'n, Poland}
\author{Franco Nori}
\email{fnori@riken.jp}
 \affiliation{Theoretical Quantum Physics
Laboratory, RIKEN Cluster for Pioneering Research, Wako-shi,
Saitama 351-0198, Japan}
\affiliation{RIKEN Center for Quantum Computing (RQC), Wakoshi, Saitama 351-0198, Japan}
\affiliation{Physics Department, The
University of Michigan, Ann Arbor, Michigan 48109-1040, USA}

\title{Continuous Dissipative Phase Transitions with or without Symmetry Breaking}

\begin{abstract}
The paradigm of second-order phase transitions (PTs) induced by spontaneous symmetry breaking (SSB) in thermal and quantum systems is a pillar of modern physics that has been fruitfully applied to out-of-equilibrium open quantum systems. 
Dissipative phase transitions (DPTs) of second order are often connected with SSB, in close analogy with well-known thermal second-order PTs in closed quantum and classical systems.
That is, a second-order DPT should disappear by preventing the occurrence of SSB. 
Here, we prove this statement to be wrong, showing that, surprisingly, SSB is not a necessary condition for
the occurrence of second-order DPTs in \textit{out-of-equilibrium open quantum systems}. 
We analytically prove this result using the Liouvillian theory of dissipative phase transitions, and demonstrate this anomalous transition in a paradigmatic laser model, where we can arbitrarily remove SSB while retaining criticality, and on a $Z_2$-symmetric model of a two-photon Kerr resonator. 
This new type of phase transition cannot be interpreted as a ``semiclassical'' bifurcation, because, after the DPT, the system steady state remains unique.
 \end{abstract}

\date{\today}

\maketitle

\section{Introduction}

The similarities and differences between quantum (or thermal) phase transitions (PTs) and dissipative phase transitions (DPTs) in open quantum systems are the subject of intense research~\cite{KesslerPRA12, JingPRL14, MarinoPRL2016, RotaPRL19,PRLLieu20, SorientePRR21, ArkadevNatComm21, Rossiniarxiv21}.
Criticality and critical phenomena (e.g., hysteresis~\cite{CasteelsPRA16, RodriguezPRL17, LandaPRL20,LandaPRB20} and slowing-down~\cite{CasteelsPRA17-2, FinkPRX17, FinkNatPhys18}) have been predicted, observed, and characterized for first-order DPTs. 
Central to the characterization of second-order PTs is the role of spontaneous symmetry breaking (SSB): nonanaliticity can occur when a system symmetry is
``broken'', i.e., the emergence of several steady (or ground) states that are not invariant anymore under the action of a given symmetry group \cite{SachdevBOOKPhase,Landau1936,MingantPRA18_Spectral}. 
SSB in open systems has been discussed in, e.g., Refs.~\cite{LeePRL13,BartoloPRA16,KepesidisNJP16,SavonaPRA17,RotaPRB17, BiellaPRA17,HuberPRA19,Munozarxiv20,TakemuraJOSAB21}.
The relation between criticality, symmetries, and exotic effects have also been discussed for a wide range of models~\cite{BartoloEPJST17, RotaNJP18,MunozPRA19,TindallNJP20,MingantiarXiv20,PRLLieu20}.

In this article, we analytically prove that \textit{second-order DPTs in open quantum systems can occur with or without symmetry breaking}. 
Similarly to other examples, where the phases of dissipative systems possess features which have no analogue in closed and thermal systems \cite{JinPRL13,LeePRL13,JinPRX16,DiehlNATPH2008,VerstraeteNATPH2009}, this feature can be explained by the spectral properties of a Liouvillian superoperator (i.e., the generator of the dynamics of an out-of-equilibrium open quantum system)~\cite{KesslerPRA12,MingantPRA18_Spectral,MingantiPRA19}.

We demonstrate our results with two examples. 
First, we consider a lasing model characterized by $U(1)$-SSB with a second-order DPT. 
By adding dephasing, the $U(1)$ symmetry of the model is maintained, but its phase coherence is destroyed, thus preventing SSB. 
Yet, a second-order DPT takes place.
Second, we consider the $Z_2$ symmetry breaking in parametric down-conversion.
In this case, the addition of a parity dissipator enables a DPT without $Z_2$-SSB.

Similar phenomena of continuous PTs without symmetry breaking (whose explanation goes beyond the Landau theory of PTs
\cite{Landau1936,Landau_BOOK_Statistical,Landau_collection1965}), can be encountered in closed systems at {\it equilibrium}~\cite{Kitaev2009,Chen2013}. These examples are characterized by a nontrivial topological structure, e.g., topological insulators~\cite{Hasan2010,Qi2011}. As such, our work
prompts the question for a generalization of topological PTs in {\it non-equilibrium}, non-quadratic, and bosonic systems~\cite{Bardyn2013,Gong2018,LieuPRL20}.
Other models of open quantum systems can undergo critical phenomena without a trivial SSB.
For instance, in Ref.~\cite{HannukainenPRA18}, it was shown that a continuous phase transition without symmetry breaking can take place.
However, such a model can also be associated with the emergence of boundary time crystals \cite{IeminiPRL18} (signaling its profoundly dissipative nature), and meaning that a time-invariance symmetry is broken.
In other words, in this model there exist multiple (oscillating) non-decaying states. 

Furthermore, the presence of a DPT without SSB has been discussed for chain of S-level spins in Ref.~\cite{HuberPRA20}.
The authors demonstrate that in a regime where the semiclassical solution of a $U(1)$ symmetric model would predict SSB, purely fluctuation-induced suppression of symmetry breaking can prevent states from retaining their coherence.

Our work reveals that in \textit{any} dissipative continuous second-order phase transition, SSB can be arbitrarily removed.
This \textit{always} allows obtaining a unique steady state in the ``broken symmetry'' region. 
As we also discuss below, such nontrivial effect can be obtained both with the use of Hamiltonian or dissipative terms, showing the nontrivial interplay between quantum and dissipative fluctuations.

\section{Criticality of open quantum systems}

Under the Born and Markov approximations \cite{BreuerBookOpen}, the reduced density matrix $\rhot$ of an open quantum system at
time $t$ evolves according to a Lindblad master equation ($\hbar=1$):
\begin{equation}\label{Eq:Lindblad1}
\frac{\de}{\de t}\rhot={\cal L}\hat\rho(t)=-i\left[\hat
H,\rhot\right] +\sum\limits_{j} \DD[\hat{L}_j]\rhot,
\end{equation}
where $\hat{H}$ is the Hamiltonian describing the coherent part of the system evolution, $\LL$ is the Liouvillian superoperator~\cite{Carmichael_BOOK_2,LidarLectureNotes,MingantiPRA19,BreuerBookOpen}, and $\DD[L_j]$ are the so-called Lindblad dissipators, whose action is
\begin{equation}\label{Eq:Dissipator_superoperator}
    \DD[\hat{L}_j]\rhot=\hat L_i\rhot\hat L_j^{\dagger} - \frac{\hat L_j^{\dagger}\hat L_j\rhot+\rhot\hat L_j^{\dagger}\hat L_j}{2}.
\end{equation}
The operators $\hat{L}_j$ are the jump operators, and they describe how the environment acts on the system inducing loss and gain of particles, energy, and information. 
The steady state $\sss$, i.e., the state which does not evolve anymore under the action of the Liouvillian ($ \partial_t \sss=\LL \sss = 0$), is central to a system characterization. We indicate the expectation values of operators at the steady state as $\expec{\hat{o}}_{\rm ss} =\operatorname{Tr}[\sss \hat{o}]$.

A DPT is a discontinuous change in $\sss$ as a function of a single parameter~\cite{MingantPRA18_Spectral,KesslerPRA12}. 
For instance, the medium gain rate $A$, defined in \eqref{Eq:L123}, plays this role in the first model considered below, and:
\begin{equation}\label{Eq:definition_PT}
    \lim_{A \to A_c} \frac{\partial^{2}}{\partial{A}^{2}} \sss(A)  \to \infty,
\end{equation}
where $A_c$ is the critical point.

This thermodynamic nonanalyticity can be witnessed in finite-size systems, as discussed in Refs.~\cite{KesslerPRA12, MingantPRA18_Spectral} and experimentally demonstrated in Refs.~\cite{RodriguezPRL17, FitzpatrickPRX17,FinkNatPhys18}.
Criticality is accompanied by the emergence of the so-called \textit{critical slowing down}, i.e., the appearance of infinitely-long timescales in the system dynamics. 
Diverging timescales can be captured by the Liouvillian spectrum, defined by
\begin{equation}\label{Eq:Definition_spectrum}
\LL \eig{i}=\lambda_i \eig{i},
\end{equation}
$\lambda_i$ (the eigenvalues) representing the decay rates and oscillation frequencies, and $\eig{i}$ (the eigenmatrices)
encoding the states explored along the dynamics of $\LL$.

\section{Dissipative phase transitions with or without spontaneous symmetry breaking}

Before dealing with a specific model, let us provide a demonstration of this novel type of criticality.

The weak symmetry of a dissipative system can be described by a superoperator $\mathcal{U}$ such that $\mathcal{U} = \hat{J} \cdot \hat{J}^{\dagger} $, where $\hat{J}^\dagger = \hat{J}^{-1}$ \cite{BaumgartnerJPA08}.
This is always the case if we assume that $\mathcal{U}$ defines a cyclic group (such as $Z_n$) and therefore $\hat{J}^\dagger \hat{J} = \hat{1}$.
Note that this is one of the most common types of symmetries that one encounters in open quantum systems, characterizing, e.g., lasing ($U_1$), parametric downconversion ($Z_2$), and $Z_n$ for a translational invariant lattice with $n$ sites.

The fact that the Liouvillian is symmetric, i.e., $[\mathcal{L}, \mathcal{U}]=0$ allows partitioning the space in different \emph{symmetry
sectors}, i.e.,  parts of the Liouvillian space which are not connected to other parts (sectors) by the system dynamics~\cite{MoodiePRA18,MingantiarXiv20,PalacinoPRR21}.
Accordingly, the Liouvillian reads $\LL = \LL_{0}+ \LL_{1} + \dots$, where $\LL_{k}$ is the evolution operator of the $k$-th symmetry sector.
As such, we can relabel the eigenvalues and eigenmatrices as $\lambda_i^{(k)}$ and $\eig{i}^{(k)}$, respectively, and write
\begin{equation}\label{Usec}
    \LL_{k} \eig{j}^{(k)}= \lambda_i^{(k)} \eig{j}^{(k)}, \,\,\, \text{ and } \,\,\,
    \mathcal{U} \eig{j}^{(k)} = \hat{J} \eig{j}^{(k)} \hat{J}^\dagger = u^{(k)} \eig{j}^{(k)}.
\end{equation}
We order the eigenvalues of each symmetry sector in such a way that
$\left|\Re{\lambda_0^{(k)}}\right| \leq \left|\Re{\lambda_1^{(k)}}\right| \leq \left|\Re{\lambda_2^{(k)}}\right|  \leq \dots $.
In this regard, $\lambda_0^{(k)}$ represents the slowest-decaying process in each symmetry sector.
The steady state must always be such that $\mathcal{U}\sss =\sss$ \cite{MingantPRA18_Spectral}.
We call this the symmetry sector for $k=0$, and therefore $\eig{0}^{(0)}\propto \sss$.
Thus, each eigenmatrix belonging to the  symmetry sector for $k=0$, i.e., $\eig{j}^{(0)}$, is such that $\mathcal{U} \eig{j}^{(0)}=\eig{j}^{(0)}$.

These considerations on symmetries are valid independently of the presence of a DPT, but they are fundamentally related to SSB.
{Generally speaking, SSB occurs when a state does not have the same symmetries, across a phase transition, as the theory that describes
it.
Such a transition is characterized by an order parameter, i.e., a quantity distinguishing phases. SSB is then signalled by the fact that the order parameter is zero (due to symmetry) in one phase  and in another phase it is non-zero ~\cite{Beekman2019}
In an open quantum system, SSB means that several steady states emerge in a region \textit{after} a DPT took place, and each of these states is not an eigenstate of $\mathcal{U}$.
}

Consider now a Liouvillian that, in a certain region of parameter space, admits a unique eigenvalue $\lambda_0^{(0)}=0$  and all the other eigenvalues are such that $\lambda_0^{(j)}\neq 0$ (i.e., $\sss \propto \eig{0}^{(0)}$, such that $\LL \sss=0$, is the only eigenmatrix which does not evolve under the Lindblad master equation).
Within this formalism, a phase transition with SSB means that in each symmetry sector a zero eigenvalue $\lambda_{0}^{(k)}$ emerges, i.e.,
\begin{equation}\label{Eq:SSB_Liouvillian_def}
    \LL \eig{0}^{(k)} \neq 0 \text{ if } { A<A_c}, \,\,\, \text{ and }  \,\,\, \LL \eig{0}^{(k)} = 0 \text{ if } { A>A_c},
\end{equation}
where the critical point $A_c$ is defined in \eqref{Eq:definition_PT}.
{Indeed, \eqref{Eq:SSB_Liouvillian_def} states that for $A>A_c$ it is possible to construct new steady states (i.e., well-defined density matrices which do not evolve under the action of $\LL$) of the form $\sss^{(k)} = \eig{0}^{(0)}+ c_k [\eig{0}^{(k)} +(\eig{0}^{(k)})^\dagger]$, and, obviously, $\mathcal{U} \sss^{(k)} \neq \sss^{(k)}$.
The role of the order parameters is then played by an operator $\hat{O}$ such that
$\operatorname{Tr}[\eig{0}^{(0)} \hat{O}]=0$, and $\operatorname{Tr}[\eig{0}^{(k)} \hat{O}]\neq 0$
Thus, \eqref{Eq:SSB_Liouvillian_def} reveals a DPT with SSB.
}

\subsection{Preventing the spontaneous-symmetry breaking}

Let us now consider a new Liouvillian
\begin{equation}
    \LL' = \LL +\DD[\hat L].
\end{equation} 
If $\LL$ is a well-defined Liouvillian, so does $\LL'$, because adding a dissipator to a Liouvillian keeps $\LL'$ a completely positive and trace-preserving (CPTP) map.
Note that the following property holds
\begin{equation}
    \DD[\hat L] \eig{j}^{(k)} = \hat L \eig{j}^{(k)} \hat L^\dagger - \frac{\hat L^\dagger \hat L \eig{j}^{(k)}+ \eig{j}^{(k)} \hat L^\dagger \hat L}{2}  \neq 0 ,  \,\,\, \text{ if and only if } \,\,\, k\neq 0,
\end{equation}
if
\begin{equation}\label{Eq:commutation_unaffected}
    \left[\eig{j}^{(0)},  \hat L\right] = \left[\eig{j}^{(0)},  \hat L^\dagger \right] = 0, \,\,\, { \text {and} } \,\,\, \left[\eig{j}^{(k)},  \hat L\right] = \left[\eig{j}^{(k)},  \hat L^\dagger \right] \neq 0.
\end{equation}
All  $\eig{j}^{(0)}$ must remain unchanged, while for $k\neq 0$, the dynamics of $\LL$ and of $\LL'$ must differ.
But since $\lambda_j^{(0)}\leq0$, and the dynamics must be different, it can only occur that $\lambda_j^{(0)}\neq 0$.

This operator can \textit{always} be found.
Indeed, we can choose  the symmetry operator $\hat{J}$ as a jump operator because
\begin{equation}\label{Eq:symmetry_unaffected}
    \DD[\hat{J}] \eig{j}^{(k)} = \hat{J} \eig{j}^{(k)} \hat{J}^\dagger - \frac{\hat{J}^\dagger \hat{J} \eig{j}^{(k)}+ \eig{j}^{(k)} \hat{J}^\dagger \hat{J}}{2} = \mathcal{U} \eig{j}^{(k)}  - \eig{j}^{(k)} = \left(u^{(k)}-1 \right) \eig{j}^{(k)}=0, \quad {\text{if and only if}} \quad k=0,
\end{equation}
according to \eqref{Usec}, and given $\hat{J}^\dagger \hat{J}=\hat{1}$.
We conclude that \textit{one can arbitrarily remove SSB from any second-order DPT}.
This does not mean that $\hat{J}$ is the only operator which allows removing the SSB while keeping a second-order DPT. It can also be done, e.g, by a generator of a symmetry group $\hat J$, provided that $\hat J$ is continuous,  as we show below on the example of a $U(1)$ model.

A remark is in order. 
If we were to include an additional term in the Hermitian Hamiltonian, not a Lindbladian dissipator, in \eqref{Eq:Lindblad1},
the above reasoning would remain the same. 
Indeed, by considering $\hat{H}'= \hat{H}+\hat{L}$ (where now $\hat{L}$ needs to be Hermitian), the commutator  $[\eig{j}^{(0)}, \hat L]=0$  ensures that the dynamics is again unchanged for the sector $k=0$.

\subsection{Second-order phase transitions and spontaneous-symmetry breaking in the Landau theory}

Phase transitions do not necessarily imply an SSB. 
For instance, a first-order phase transition can occur even for non-symmetric systems \cite{Landau_BOOK_Statistical} (this is true also in out-of-equilibrium systems \cite{VicentiniPRA18,MingantPRA18_Spectral,Foss-FeigPRA17,CasteelsPRA17-2,BiondiPRA17}).
There are also examples of second-order phase transitions without SSB, such as topological phase transitions, including the Berezinskii–Kosterlitz–Thouless transition in two-dimensional XY model \cite{KosterlitzJPC73} or Bose gasses \cite{HadzibabicNat06}. In this case, however, topology dictates the change in the system properties.

The phenomenological theory of phase transitions, developed by Landau \cite{Landau_BOOK_Statistical}, explains second-order criticality in terms of symmetries and SSB.
One phase ``possesses'' a certain symmetry (i.e., the system state is invariant under the action of the symmetry group), while, in the SSB, phase this symmetry property is no more exhibited.
One assumes that there is a thermodynamic potential $F$ (e.g., a free-energy function) that is an analytic function of an order parameter $\beta$.
The presence of a symmetry constrains the form of the potential which, up to a constant shift, will read
\begin{equation}
F =A \beta^2 +B \frac{\beta^4}{4}.
\end{equation}
The order parameter $\beta$ will be the one which minimizes the potential $F$.
While $A$ and $B$ explicitly depend on the various parameters (such as temperature, pressure, chemical potential, etc.), thermodynamic considerations on the system stability imply that $B>0$.
Since $A>0$ implies that $\beta=0$ minimizes $F$, a second-order DPT occurs when $A$ becomes negative.
At this point, the order parameter $\beta$ that minimizes $F$ is no more nonzero, and therefore a continuous, but nonanalytical, change in the system parameters occurs. 
Since symmetry dictates that $\beta=0$, such a second-order phase transition is always connected to a SSB.
The Landau theory remains applicable also for the description of the continuous {\it quantum} phase transition with SSB,  where one can construct the free energy functional from a given microscopic Hamiltonian~\cite{SachdevBOOKPhase}. 

Our result, instead, indicates that SSB can be arbitrary removed from second-order DPTs:
a Landau-like theory cannot correctly capture the nature of DPTs.

\color{black}

\section{Model I: A $U(1)$-symmetric laser model}

\begin{figure}
    \centering
    \includegraphics[width=0.96 \textwidth]{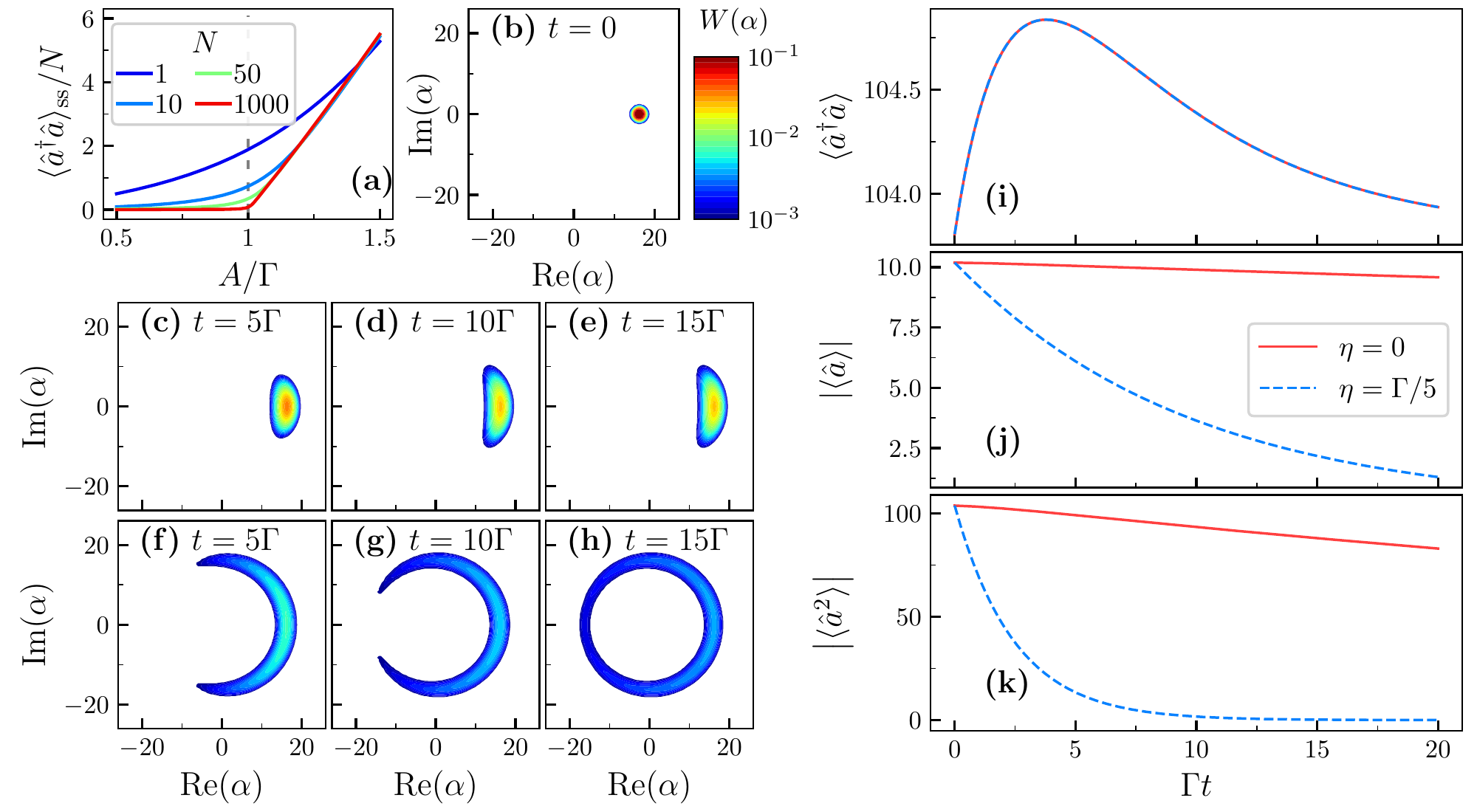}
    \caption{Dissipative phase transition with or without the $U(1)$ spontaneous symmetry breaking.
    (a) Rescaled number of photons $\expec{\hat{a}^\dagger \hat{a}}_{\rm ss}/N$ versus incoherent drive strength $A/\Gamma$ for various $N$. The solid curves are the results of the quantum simulations, and are independent of $\eta$. 
    (b) Wigner function of a state initialized in the coherent state $\ket{\alpha=\sqrt{\expec{\hat{a}^\dagger \hat{a}}_{\rm ss}}}$ for $A=1.25\Gamma$ and $N=50$ evolving with:
    (c-e) $\eta=0$; (f-h) $\eta=\Gamma/5$. If $\eta=0$, the dissipative phase transition coincides with a spontaneous symmetry breaking (shown by the long-time coherence).
    If $\eta\neq0$, a dissipative phase transition occurs because $\expec{\hat{a}^\dagger \hat{a}}_{\rm ss}/N$ is discontinuous, but the system has no spontaneous symmetry breaking since it rapidly loses coherence.
    The time evolution of the operators $\expec{\hat{a}^\dagger \hat{a}}$, $|\expec{ \hat{a}}|$, and $|\expec{ \hat{a}^2}|$ for $\eta=0$ (solid red curve) and $\eta=\Gamma/5$ is shown in (h-k), respectively.
    Parameters: $B/\Gamma=10^{-1}/N$ and $\omega/\Gamma=0$ (i.e., the frame rotates at $\omega$).
    }
    \label{fig:number_wigner}
\end{figure}

Let us provide an example of a DPT where the SSB can be arbitrarily removed. Consider a laser-like $U(1)$ model with $\hat{H}=\omega
\hat{a}^\dagger \hat{a}$ and jump operators
\begin{equation}\label{Eq:L123}
\begin{split}
\hat L_1 =\frac{\ap(2A- B \am\ap)}{2\sqrt{A}}, \,\,\, \hat L_2 =
\sqrt{\frac{3B}{4}}\,\am\ap, \,\,\, \hat L_3 = \sqrt{\Gamma}\am,
\end{split}
\end{equation}
where $\hat{a}$ ($\hat{a}^\dagger$) is the bosonic annihilation (creation) operator, $\hat{L}_1$ describes the laser gain, $\hat{L}_2$ captures the field dephasing, and $\hat L_3$ represents the particle loss. 
The jump operators are characterized by the rates: $A$ for the medium gain, $B$ for the gain saturation, $\Gamma$ for the dissipation (the inverse of the photon lifetime)
Changing to the frame rotating at the frequency $\omega$, we can set $\hat{H}=0$ \cite{MingantiarXiv20}. This model is the celebrated Scully-Lamb laser master equation in the so-called weak-gain saturation regime \cite{YamamotoBook,ScullyLambBook,Gea1998,Arkhipov2019}, valid if
$A=\mathcal{O}(\Gamma)$ and ${B}\expec{\hat{a}\hat{a}^\dagger}\ll{2A}$.
The limits of the validity of this approximation for the system dynamics are detailed in, e.g., Refs.~\cite{WangPRA73,minganti2021liouvillian}.

The model is characterized by a $U(1)$ \textit{weak symmetry} \cite{BucaNPJ2012,AlbertPRA14}, which is represented by the symmetry operator $\hat J=\exp\left(i\phi\hat a^{\dagger}\hat a\right)$. Indeed, the transformation $\hat{a}\to \hat J\hat a\hat J^{\dagger}\to\hat{a} e^{i \phi}$ leaves the equation of motion unchanged, but $\expec{\hat{a}^\dagger \hat{a}(t)}$ is \textit{not} conserved. 
Thus, $\expec{\hat{a}}_{\rm ss} = 0$ holds for any finite-size system. A SSB in the thermodynamic limit is, thus, signalled by $\expec{\hat{a}}(t\to \infty) \neq 0$, which follows from symmetry considerations~\cite{DeGiorgio1970}. 
Therefore, in finite-size systems, the $U(1)$ SSB is  signalled by $\expec{\hat{a}}(t) \neq 0$ for increasingly long time, as we increase $N$.
To better grasp the meaning of this symmetry, let us express the eigenmatrices $\eig{i}$ of the Liouvillian and the action of the symmetry operator  in \eqref{Usec} in the number (Fock) basis:
\begin{equation}\label{rho_i}
    \eig{j}^{(k)}=\sum_{m,n} c_{m,n}^{(k)}\ket{m}\bra{n} \,\, \text{ and } \,\, \mathcal{U}\eig{j}^{(k)} = \sum_{m,n} c_{m,n}^{(k)} e^{-i \phi \hat{a}^\dagger \hat{a}}  \ket{m}\bra{n} e^{i \phi \hat{a}^\dagger \hat{a}} =\sum_{m,n} c_{m,n}^{(k)} e^{-i \phi (m-n)}\ket{m}\bra{n}= u_i^{(k)} \eig{i}^{(k)}.
\end{equation}
We conclude that $\exp\left[{-i \phi (m-n)}\right]$ must be a constant and, therefore, any eigenmatrix $\hat\rho_i$ in Eq.~(\ref{rho_i}) must obey
\begin{equation}\label{Eq:condition_symmetry}
    \eig{i}^{(k)}= \sum_{m} c_{m}^{(k)} \ket{m}\bra{m-k} \, ,
\end{equation}
for some constant integer $k\in \mathbb Z$.
In other words, $\eig{i}^{(k)}$ must be an operator containing elements only on one diagonal, and different symmetry sectors occupy different upper and lower diagonals.

\subsection{Second-order dissipative phase transitions with the spontaneous
breaking of $U(1)$}

PTs and nonanaliticity can only emerge in the thermodynamic limit. One can exploit the infinite dimension of the bosonic Hilbert space to observe a nonanalytical change in the steady state, as discussed, in, e.g. Refs.~\cite{CarmichaelPRX15,CasteelsPRA17,BartoloPRA16, MingantPRA18_Spectral, CurtisPRR21, MingantiarXiv20}. To do so, we consider a rescaling parameter $N$, so that the size of the Hilbert space increases, but a meaningful
observable, such as the rescaled photon number, merge far from the critical point:
\begin{equation}\label{Eq:Scaling}
\{A, B, \Gamma\} \to \{A, B/N, \Gamma\}.
\end{equation}
In a laser model \cite{YamamotoBook}, $N$ represents an increasing number of injected three-level atoms in the lase cavity, but each
with a weaker light-matter coupling.
We also refer to Ref.~\cite{minganti2021liouvillian} for a more detailed discussion of the system thermodynamic limit and on the nature of the DPT in the Scully-Lamb laser model.

To numerically simulate the results of this model, we introduce a cutoff $C$ in the Hilbert space, i.e., 
we assume that $\expec{m|\rhot|n}=0$ if $m>C$ or $n>C$.
We then verify the convergence with the cutoff, i.e., we check that by increasing $C$ the values do not change
(within a numerical precision).
By decreasing the nonlinearity (i.e., increasing $N$), $C$ increases.

In Fig.~\ref{fig:number_wigner}(a), we plot the rescaled photon number $\expec{\hat{a}^\dagger \hat{a}}_{\rm ss}/N$, obtained by
numerically solving $\LL \sss =0$ (solid curves), and the results of the semiclassical approximation (dashed lines) for different
values of $N$.
As one can see, the photon number is continuous, but there is an emerging ``elbow'' signalling the occurrence of a second-order DPT.

The breaking of the $U(1)$ symmetry means the retaining of coherence for infinitely long time.
Although this occurs only in the thermodynamic limit, we can show the presence of the critical slowing down for finite-size systems. 
In Figs.~\ref{fig:number_wigner}(b)-\ref{fig:number_wigner}(e), we show the Wigner function $W(\alpha)$ for an initially coherent
state when $\eta=0$ in the ``broken symmetry region'' (i.e., for $A>A_c=\Gamma$),  where
$  W(\alpha) = 2\operatorname{Tr}\left[\hat{D}_{\alpha} \exp\left(i
\pi \hat{a}^{\dagger} \hat{a}\right) \hat{D}_{\alpha}^{\dagger}
\rhot \right]/\pi$ and $\hat{D}_{\alpha}=\exp{(\alpha \hat{a}^\dagger - \alpha^{*}
\hat{a})}$ is the displacement operator~\cite{Haroche_BOOK_Quantum}. 
As time passes, the system retains its coherence $\expec{\hat{a}(t)}\neq 0$, signalling that, even for this finite-size system, a critical timescale associated with SSB has emerged.

To better quantify the meaning of SSB and the role of the $U(1)$ symmetry, we note that
\begin{equation}
    \Tr{\hat{a}^{n} \eig{j}^{(k)} } = \Tr{ \mathcal{U}^\dagger \mathcal{U} \hat{a}^{n}  \eig{j}^{(k)}} =  \Tr{\left(e^{+i \phi \hat{a}^\dagger \hat{a}} \hat{a}^{n}  e^{-i \phi \hat{a}^\dagger \hat{a}} \right) \left( e^{-i \phi \hat{a}^\dagger \hat{a}} \eig{j}^{(k)} e^{i \phi \hat{a}^\dagger \hat{a}} \right)} = e^{-i \phi (k-n)} \expec{\hat{a}^{n} \eig{j}^{(k)} }.
\end{equation}
We, thus, conclude that the critical slowing down, associated with SSB in the $k$th symmetry sector, is witnessed by $\expec{\hat{a}^{k}}$.
Thus, in Figs.~\ref{fig:number_wigner}(i-k) we plot with red solid curves the dynamics of $\expec{\hat{a}^\dagger \hat{a}(t)}$, $|\expec{ \hat{a}(t)}|$, and $|\expec{ \hat{a}^2(t)}|$ (associated with the sectors $k=0, \, \pm 1,\, \pm 2$), respectively. 
The initial state is the same coherent state as in Figs.~\ref{fig:number_wigner}(b).
While the photon number rapidly converges to its steady-state value, we see again that the coherences are preserved for very long-times, confirming that indeed a $U(1)$ SSB is taking place, and there is a critical slowing down in each symmetry sector.

\subsection{Removing the spontaneous breaking of $U(1)$ symmetry}

According to our proof, we should be able to remove the SSB while retaining criticality.
To do that, we notice that a jump operator of the form $\hat{L}=\sqrt{\eta/4} \am\ap$, where $\eta$ represents an additional dephasing rate, satisfies \eqref{Eq:commutation_unaffected} given the structure of $\eig{j}^{(k)}$ [c.f. \eqref{Eq:condition_symmetry}].

First, we verified that the photon number is identical to the one for which $\eta=0$ [the results are within a floating-point precision in Fig.~\ref{fig:number_wigner}(a)]. 
Again, the photon number becomes sharper and sharper with increasing $N$, and the results coincide for all tested values of $\eta$, meaning that a second-order DPT is taking place.

Figs.~\ref{fig:number_wigner}(f)-\ref{fig:number_wigner}(h) show
that an initially coherent state, as in the case $\eta\neq0$, rapidly looses
its coherence on a timescale of $\eta/2$. That is, the $U(1)$ SSB
does not take place. Indeed, $\rhot \simeq \sss$ in
Fig.~\ref{fig:number_wigner}(h), indicating that an initial state
rapidly reaches its steady state, proving the absence of any
residual or hidden SSB, which would anyhow lead to a critical
slowing down.

To better quantify how each symmetry sector is affected, we compare
$\expec{\hat{a}^\dagger \hat{a}(t)}$, $|\expec{ \hat{a}(t)}|$, and $|\expec{ \hat{a}^2(t)}|$ 
in Figs.~\ref{fig:number_wigner}(i-k) for $\eta=0$ (red solid curves) and $\eta\neq 0$ (blue dashed curves). 
Wile the sector for $k=0$ is unaffected (as demonstrated by the photon number). Dynamics in other symmetry sectors is much faster, and $|\expec{ \hat{a}^{n}(t)}|$ rapidly reaches zero.

We have, thus, demonstrated the existence of a second-order DPT
without the $U(1)$ SSB and the possibility to arbitrary remove the SSB
retaining a critical behavior. 
Importantly, the semiclassical analysis~\cite{DeGiorgio1970} fails to capture this kind of criticality (see also Ref.~\cite{minganti2021liouvillian} for details),
which highlights the necessity to use the Liouvillian formalism in describing dissipative critical phenomena.

\begin{figure}
    \centering
    \includegraphics[width=0.96 \textwidth]{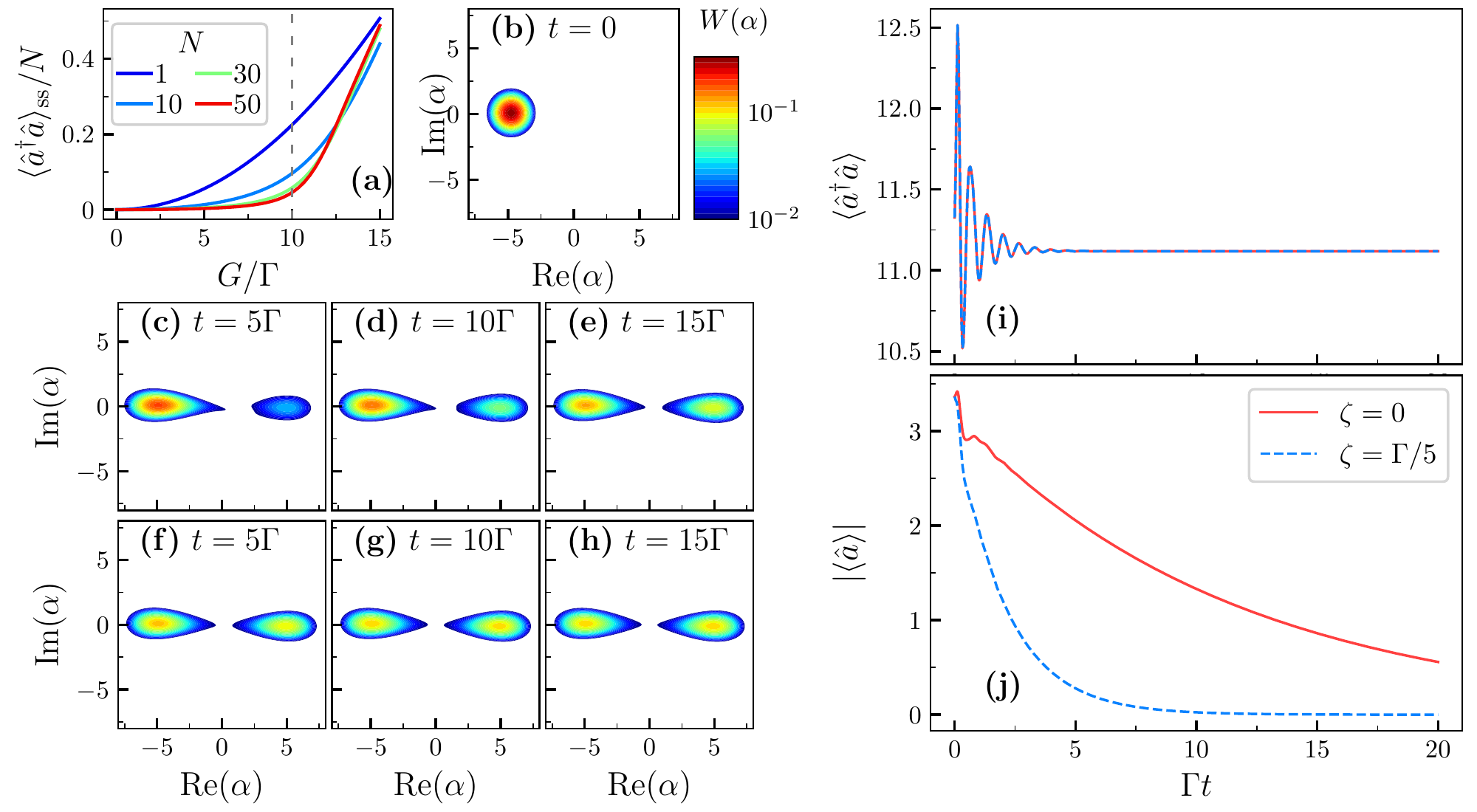}
    \caption{Dissipative phase transition with or without the $Z_2$ spontaneous symmetry breaking.
    (a) Rescaled number of photons $\expec{\hat{a}^\dagger \hat{a}}_{\rm ss}/N$ versus incoherent drive strength $A/\Gamma$ for various $N$. The solid curves are the results of the quantum simulations, and are independent of $\zeta$. 
    (b) Wigner function of a state initialized in the coherent state $\ket{\alpha=\sqrt{\expec{\hat{a}^\dagger \hat{a}}_{\rm ss}}}$ for $G=12.5\Gamma$ and $N=50$ evolving with:
    (c-e) $\zeta=0$; (f-h) $\zeta=\Gamma/5$. If $\zeta=0$, the dissipative phase transition coincides with a spontaneous symmetry breaking (shown by the long-time coherence).
    If $\eta\neq0$, a dissipative phase transition occurs because $\expec{\hat{a}^\dagger \hat{a}}_{\rm ss}/N$ is discontinuous, but the system has no spontaneous symmetry breaking since it rapidly loses coherence.
    The time evolution of the operators $\expec{\hat{a}^\dagger \hat{a}}$ and $|\expec{ \hat{a}}|$ for $\zeta=0$ (solid red curve) and $\zeta=\Gamma/5$ are shown in (i-j), respectively.
    Parameters: $\Delta/\Gamma=10$ and $U/\Gamma=10$.
    }
    \label{fig:number_wigner_Kerr}
\end{figure}

\section{Model II: A $Z_2$-symmetric Kerr resonator}

To further illustrate the validity of our results, here we consider the second-order DPT 
of a two-photon Kerr resonator, studied in, e.g., \cite{BartoloPRA16,MingantPRA18_Spectral,BartoloEPJST17,SavonaPRA17}.
The Hamiltonian in the frame rotating at the pump frequency reads
\begin{equation}
\label{KerrRes}
\hat{H} = -\Delta \hat{a}^\dagger \hat{a} +i \frac{G}{2} \left[ \left(\hat a^\dagger\right)^2 -  \hat{a}^2  \right]+ \frac{U}{2} \left(\hat a^\dagger\right)^2  \hat{a}^2 ,
\end{equation} 
where $\Delta$ is the cavity-to-pump detuning, $G$ is the two-photon drive intensity, and $U$ is the Kerr nonlinear interaction. 

Photons continuously escape the Kerr resonator, and the system is described by the Lindblad master equation for the system density matrix $\rhot$ \cite{Wiseman_BOOK_Quantum}
\begin{equation} \label{Eq:LME}
 \frac{\partial}{\partial t} \hat{\rho}(t) = -i [\hat{H}, \hat{\rho}(t)] + \Gamma \DD[\hat{a}] \rhot.
\end{equation}

This system is characterized by a $Z_2$ weak symmetry, meaning that $\hat{a}\to \hat J\hat a\hat J^{\dagger} = - \hat{a}$ leaves the equation of motion unchanged, where $\hat J=\exp\left(i\pi \hat a^{\dagger}\hat a\right)$, but parity is not a conserved quantity \cite{AlbertPRA14}. 
With a reasoning similar to that of \eqref{rho_i}, we can demonstrate that there are two symmetry sectors, $k=0$ and $k=1$, such that
\begin{equation}\label{Eq:condition_symmetry_Z2}
    \mathcal{U}\eig{i}^{(k)} = e^{i k \pi } \eig{i}^{(k)}, \, \text{ and } \,  \eig{i}^{(k)}= \sum_{m, n} c_{2m, \, 2n}^{(k)} \ket{2m}\bra{2n+k} + c_{2m+1, \, 2n+1}^{(k)} \ket{2m+1}\bra{2m+1+k} \, .
\end{equation}
We conclude that $\eig{i}^{(0)}$ contains only the even-even and odd-odd states, while $\eig{i}^{(1)}$ couples the even-odd and odd-even states.

This time, the rescaling parameter $N$ acts as 
$\{\Delta, U, G, \Gamma\} \to \{\Delta, U/N, G, \Gamma\}$ \cite{BartoloPRA16,MingantiSciRep16,DiCandiaArXvi21}.

Similarly to the previous case, we observe the emergence of a second-order DPT [Fig.~\ref{fig:number_wigner_Kerr}(a)].
This is accompanied by a critical slowing down signaling a SSB, as it can be argued from the evolution of a system initialized in a coherent state.
In particular, both the Wigner functions in Figs.~\ref{fig:number_wigner_Kerr}(b-e) and the evolution of $\expec{\hat{a}^\dagger \hat{a}}$ and $|\expec{\hat{a}}|$ [solid lines in Figs.~\ref{fig:number_wigner_Kerr}(i-j)] show the presence of a slow-time scale in the $k=1$ symmetry sector.

To remove the SSB and keep the second-order DPT, this time we consider an additional jump operator of the form $\hat{L}= \sqrt{\zeta} \hat{J}= \sqrt{\zeta} \exp\left(i\pi \hat a^{\dagger}\hat a\right)$.
Again, there is no difference with the case for $\zeta=0$ in the symmetry sector for $k=0$, as it can be argued by the fact that the photon number in the steady state is unchanged  [Fig.~\ref{fig:number_wigner_Kerr}(a)] as well as the time dynamics of $\expec{\hat{a}^\dagger \hat{a}}$ [Fig.~\ref{fig:number_wigner_Kerr}(i)].
However, the presence of $\zeta$ significantly changes the sector $k=1$, as it can be seen by analyzing the Wigner function in Figs.~\ref{fig:number_wigner_Kerr}(f-h) and the time evolution of $|\expec{\hat{a}}|$ in Fig.~\ref{fig:number_wigner_Kerr}(j).

We confirm again our predictions, and we show that, by adding an appropriate dissipator, we can remove the SSB also in the case of this $Z_2$ SSB.

\section{Conclusions}

In this article, we have 
proved that 
continuous DPTs can occur with and
without an SSB. 
In particular, we derived analytical conditions to remove SSB from any second-order DPT.
As examples, we have analyzed a paradigmatic
non-equilibrium lasing system and a model characterized by a discrete $Z_2$ symmetry.
In both cases, our analytical predictions 
are confirmed by the numerical simulations,
demonstrating how tremendously multiform and various are 
the non-equilibrium states, their dynamics, and their phase transitions.

The presented DPTs without SSB are, to our knowledge, a novel
phenomenon, where dissipation plays a fundamental role.
This phenomenon opens questions concerning the mechanism of criticality in open quantum systems.
Our predictions can be experimentally
tested with, e.g., superconducting circuits, where physically engineered
dissipation can be realized with state-of-the-art techniques. 

From a fundamental point of view, the presence of second-order DPTs, where degeneracy can be removed, is intriguing because
it represents a shift from the Landau theory of phase transitions. 
An interesting question would now be if it is possible, with similar mechanisms, to ``reintroduce'' SSB in second-order DPTs without SSB, such as that in Ref.~\cite{HuberPRA20}.
Revealing a link, if any, with extensions of topological
theories for DPTs is one of our future objectives
\cite{LieuPRL20,Gong2018}.
Even though the models considered here were characterized by a single order parameter, our results are valid, in general, 
for any multi-order parameter systems with any global symmetry, 
e.g., open and/or dissipative extensions of thermal ${\mathbb Z}_2$-symmetric systems 
with light-matter interactions~\cite{Schiro2016,Bazhenov2021}.

\begin{acknowledgments} 
The authors are grateful to the RIKEN Advanced
Center for Computing and Communication (ACCC) for the allocation
of computational resources of the RIKEN supercomputer system
(HOKUSAI BigWaterfall). I.A. thanks the Grant Agency of the Czech
Republic (Project No.~18-08874S), and Project no.
CZ.02.1.01\/0.0\/0.0\/16\_019\/0000754 of the Ministry of
Education, Youth and Sports of the Czech Republic. A.M. is
supported by the Polish National Science Centre (NCN) under the
Maestro Grant No. DEC-2019/34/A/ST2/00081. F.N. is supported in
part by:
 Nippon Telegraph and Telephone Corporation (NTT) Research,
 the Japan Science and Technology Agency (JST) [via
 the Quantum Leap Flagship Program (Q-LEAP) program,
 the Moonshot R\&D Grant Number JPMJMS2061, and
 the Centers of Research Excellence in Science and Technology (CREST) Grant No. JPMJCR1676],
 the Japan Society for the Promotion of Science (JSPS)
 [via the Grants-in-Aid for Scientific Research (KAKENHI) Grant No. JP20H00134 and the
 JSPS~CRFBR Grant No. JPJSBP120194828],
 the Army Research Office (ARO) (Grant No. W911NF-18-1-0358),
 the Asian Office of Aerospace Research and Development (AOARD) (via Grant No. FA2386-20-1-4069), and
 the Foundational Questions Institute Fund (FQXi) via Grant No. FQXi-IAF19-06.
\end{acknowledgments}


%

\end{document}